\documentclass[showpacs,twocolumn,prb,aps,color,amssymb]{revtex4}
\usepackage{graphicx}
\begin{document}

\title{Antiferromagnetic Heisenberg ladders in a staggered magnetic field}

\author{Jize Zhao$^2$, Xiaoqun Wang$^{1,2}$
%\footnote{Corresponding author. E-mail address: xiaoqunwang@ruc.edu.cn},
Tao Xiang$^{2}$, Zhaobin Su$^{2}$, and Lu Yu$^{2}$}
\affiliation{$^1$Department of Physics, Renmin University of
China, Beijing 100872, China}
 \affiliation{$^2$Institute of Theoretical Physics and Interdisciplinary Center
 of Theoretical Studies, CAS, Beijing 100080, China}
\author{Jizhong Lou and Changfeng Chen}
\affiliation{Department of Physics, University of Nevada, Las Vegas,
Nevada 89154, USA}
\date{\today}
\begin{abstract}
We study the low-energy excitations of the spin-1/2 antiferromagnetic Heisenberg chain
and $N$-leg ($N$=2, 3, 4) ladders in a staggered magnetic field $h_s$.  We show that $h_s$
induces gap and midgap states in all the cases and examine their field scaling behavior.
A modified boundary scheme is devised to extract accurate bulk excitation behavior.  The
gap values converge rapidly as $N$ increases, leading to a field scaling exponent
$\gamma=1/2$ for both the longitudinal and transverse gaps of the square lattice
($N \rightarrow \infty$).  The midgap states induced by the boundary edge effects share
the bulk gap scaling exponents but their overall scaling behavior in the
large-$N$ limit needs further investigation.
\end{abstract}


\maketitle

Heisenberg spin ladders have attracted considerable interest for
their fascinating properties due to strong quantum fluctuations
and their unique structural character as a crossover platform from
one to two dimensions.\cite{Dagotto} It is known that spin ladders
with odd and even numbered legs have drastically different
behavior; the former have gapless spin excitation spectra while
the latter have finite spin gaps.\cite{Noack,Wang00,Wang02}  This
disparity complicates considerably attempts to extrapolate the
results of ladders to two-dimensional systems.  An important
development in the study of low-dimensional spin systems is the
observation of an unexpected  magnetic field induced gap in the
low-energy excitation spectrum of copper benzoate, a
quasi-one-dimensional spin chain material.\cite{Dender} Recently,
the field induced gap has been reported in more spin chain
materials such as Yb$_4$As$_3$,\cite{Kohgi,Fulde,Osh2}
CuCl$_2$¡¤2(dimethylsulfoxide) \cite{Ken04}, and
[Pyrimidine-Cu(NO$_3$)$_2$(H$_2$O )$_2$]$_n$(CuPM).\cite{Zvy04}

A gap in the low-energy excitation spectrum of a spin-1/2
Heisenberg chain can be induced by an effective staggered magnetic
field.\cite{aderea}  The staggered field may originate from the
staggered gyromagnetic tensor or/and the Dzyaloshinskii-Moriya
(DM) interaction when an external magnetic field is
applied.\cite{Oshikawa1} Using a local gauge transformation and
neglecting small anisotropic terms in both the Heisenberg exchange
and the Zeemann splitting terms, an effective Hamiltonian for spin
chain materials with the DM interaction can be written as
\cite{Oshikawa1}
\begin{equation}
\hat{H}_{eff}=\sum_{i}\left[ J\hat{\bf S}_{i}\cdot\hat{\bf S}_{i+1}-HS^{z}_{i}
   -h_{s}(-1)^{i}S^{x}_{i}\right],
\label{HEFF}
\end{equation}
where $H$ and $h_s$ are the uniform and staggered magnetic field,
respectively. This effective Hamiltonian has been mapped onto the
sine-Gordon model using the bosonization technique
\cite{Oshikawa1} to obtain an analytic form for the spin gap as a
function of the magnetic field.  Furthermore, it was shown
\cite{Oshikawa1} that the uniform field does not change the
qualitative scaling behavior induced by the staggered field and
the scaling function derived for $H$=0 provides a good explanation
for the experimental results at the low-field region and its
validity is supported by the density matrix renormalization group
(DMRG) calculations.\cite{lou,zhao}  However, the effective field
theory is only applicable when the field is not too strong. The
recent DMRG calculations unveiled different field-dependence at
higher fields and a crossover in the intermediate field
regime.\cite{lou,zhao}

Additional intriguing phenomena induced by the staggered field have been recently
reported for quasi-one-dimensional spin chain materials $BaCu_2Si_2O_7$ \cite{Tsukada}
and $CuCl_{2}\cdot{2(dimethylsulfoxide)}~(CDC)$.\cite{Ken04}  The neutron
scattering measurement on $CDC$ and electron spin resonance measurement on
$BaCu_2Si_2O_7$ show that the field dependence of the induced gap deviates
from the sine-Gordon model prediction for the spin chains.  It suggests that
the interchain interaction may play an important role.  Studies on the effect of
the interchain interaction is therefore needed to clarify the fundamental physics
and to establish the nature of the low-energy excitation in these new materials.
Moreover, there is also considerable interest in searching for a reliable
extrapolation from multi-leg ladders to the square lattice.\cite{Sato,Mila}
This should be achievable for spin chain materials with the DM interaction since,
unlike the $h_s$=0 case, ladders
with both odd and even numbered legs show field induced gaps with similar scaling
behavior under the staggered field.

In the present work, we study spin-1/2 antiferromagnetic
Heisenberg ladders with the following Hamiltonian:
\begin{eqnarray}
\hat{H} & = &\sum_{a=1}^{N}\sum_{i=1}^{L-1}J\hat{\bf S}_{a,i}\cdot\hat{\bf S}_{a,i+1}
 + \sum_{a=1}^{N-1}\sum_{i=1}^{L}J_{\perp}\hat{\bf S}_{a,i}\cdot\hat{\bf S}_{a+1,i} \nonumber \\
&& + \sum_{a=1}^{N}\sum_{i=1}^{L}(-1)^{i+a}h_sS^{z}_{a,i},
\label{HLADDER}
\end{eqnarray}
where $N$ ($L$) is the number of legs (rungs).  We consider the
case of isotropic coupling, i.e.,  $J=J_{\perp}=1$ and employ the
DMRG method \cite{White1,peschel,Uli} to study its low-energy
properties. We kept up to 800 states for ladders with up to 300
rungs in our computations.  The truncation errors are less than
 $10^{-8}$ in all the cases.
Although the staggered field breaks the SO(3) symmetry, the total $S^z$ remains conserved.
The spin gaps for the longitudinal and transverse branches, $\Delta_{\mathcal{L}}$
and $\Delta_{\mathcal{T}}$, are defined as:
\begin{eqnarray}
\Delta_{\mathcal{L}}(L)=E_{1}(L,0)-E_{0}(L,0)\\
\Delta_{\mathcal{T}}(L)=E_{0}(L,1)-E_{0}(L,0)
\end{eqnarray}
where $E_{0}(L,S_{z})$ and $E_{1}(L,S_{z})$ are the ground-state and first
excitation energy in the $S_{z}$ sector.  In DMRG calculations, numerical
accuracy is usually much higher with the use of the open boundary condition (OBC)
than that with the periodic boundary condition (PBC).  Recent studies \cite{lou2}
reveal that there generally exist midgap states
induced by the edge excitations in open-end spin chains in the presence of a
staggered magnetic field.  Our calculations show that midgap states also exist in
spin ladders (see below).  This complicates the process of extracting the bulk excitations.
To address this issue, we devised a modified boundary scheme (MBS) to first identify and
then move the midgap states away from the low-energy spectrum.  It ensures a reliable
extraction of the bulk excitation gap while allowing an accurate description
of the midgap states.

The idea behind the MBS is to push the edge excitations up in energy and leave only
the bulk excitations in the low-energy spectrum.  This is achieved by introducing an
edge parameter in the Hamiltonian that systematically drives up the edge excitation.
Similar treatments have been proposed in previous DMRG calculations with
OBC.\cite{Vekic,Schollwock,wang}  Here we introduce a continuous monotonic function
$f(x),0\le{x}\le{1}$, where $f(0)>1$ and $f(1)=1$, to rescale the Hamiltonian parameters
\begin{eqnarray*}
Y_{i}/Y=\left\{
    \begin{array}{ll}
       c_{i},&1\le{i}\le{M}\\
       1, &M<i<L-M\\
       c_{L+1-i},&L-M+1\le{i}\le{L}
    \end{array},
  \right.
\end{eqnarray*}
where $Y$ represents $J$, $h_s$, or $J_{\perp}$ and $c_{m}=f(m/M),1\le{m}\le{M}$
where $M$ is the number of the edge sites on which the parameters are adjusted.
We choose the functional form $f(x)=1+\alpha(1+\cos(\pi{x}))$ where $\alpha\ge{0}$
is the only adjustable parameter ($\alpha=0$ corresponds to OBC).  This choice is
certainly not unique, but it satisfies the requirement for simplicity and effectiveness
in removing edge excitations from the low-energy spectrum.  We have run extensive tests
and found this scheme work very well for ladders with both ferromagnetic and
antiferromagnetic couplings at all field strengths.

\begin{figure}[h!]
\includegraphics[width=6.8cm,angle=0]{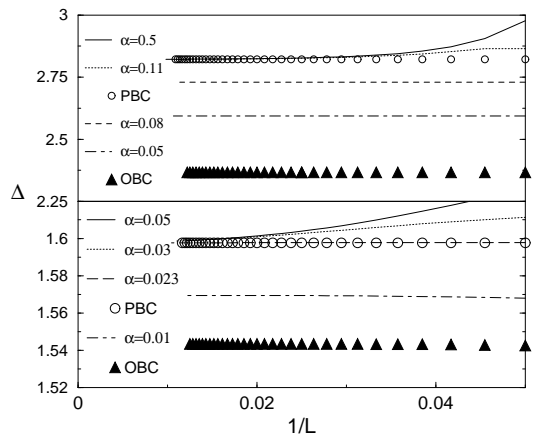}
\vspace{-0.4cm} \caption{The longitudinal (top) and transverse
(bottom) gap and midgap states versus $1/L$ at various $\alpha$
values for the two-leg ladder at $h_s$=0.6 with $M=10$.}
\label{fig1}
\end{figure}
 We demonstrate the implementation of MBS in the case of
the two-leg ladder. Figure \ref{fig1} shows the gap and midgap
states versus $1/L$ with various $\alpha$ at $h_s=0.6$.  The gap
obtained under PBC represents the bulk value. A midgap state
appears in the open-end (OBC) case due to topological edge
effects.\cite{Ng,Qin,to} As $\alpha$ increases, the midgap state
is gradually pushed up in energy and eventually enters the
continuum spectrum, leaving the bulk gap state as the lowest
excitation in the large-$L$ limit.  We performed the same
calculations on the three- and four-leg ladders and found that
midgap states exist for all the cases.  With MBS one can clearly
identify these bound states inside the field induced gap for a
detailed study and then systematically push them up in energy and
extract the bulk excitations with proper values of $\alpha$. It is
noted that a large $\alpha$ is needed for cases where the edge
state lies well below the bulk gap.  In some special cases where
the edge states are degenerate in the large-$L$ limit with the
ground state as in spin-1 and 2 Heisenberg chains, special
measures \cite{Schollwock,wang} need to be taken to eliminate the
degeneracy before applying MBS.  In the calculations presented
below we use $\alpha=2.0$ with $M=10$  to ensure good convergence
in all the cases.

\begin{figure}[h!]
\begin{center}
\includegraphics[width=6.8cm,angle=0]{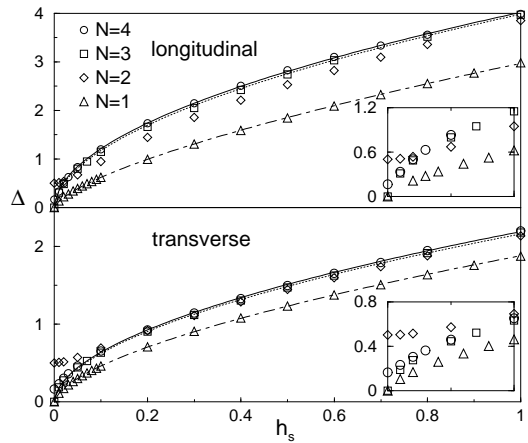}
\vspace{-0.5cm}
 \caption{The longitudinal and
transverse gap versus the staggered magnetic field $h_s$ for the
spin-1/2 chain ($N$=1) and $N$-leg ($N=2, 3, 4$) ladders.  The
fitted scaling curves are also shown for $N=1,3,4$ as the dashed,
dotted and solid lines, respectively. The insets illustrate the
gaps for $h_s\in[0,0.1]$. } \label{fig2}
\end{center}
\end{figure}

 We now turn to the study of the staggered field
induced gaps in the spin chain and ladders and the extrapolation
toward the square lattice.  Figure \ref{fig2} shows the
longitudinal and transverse gaps for the chain ($N$=1) and $N$-leg
($N$=2, 3, 4) ladders. When $h_{s}=0$ the spin excitations for
$N$=1 and 3 cases are gapless. For the two- and four-leg ladders,
the zero (staggered) field gaps are 0.502$J$ and 0.165$J$,
respectively. These are in agreement with previous DMRG results
\cite{Noack} and Monte Carlo data.\cite{Greven} For the four-leg
ladder, the present gap value is expected to be much more accurate
than those previously reported since we have kept a very large
number (up to 800) of states in the DMRG calculations.  In the
weak field limit, the gaps for the ladders with even-numbered legs
show a different field scaling behavior from that for the ladders
with odd-numbered legs (see the insets in Fig.\ref{fig2}). In
particular, it has been shown \cite{Oshikawa1,lou,zhao} that the
field induced gap for the spin-1/2 chain scales as $h_s^{\gamma}$
with a scaling exponent $\gamma=2/3$ and an additional logarithmic
correction for small $h_s$.  Meanwhile, the gaps converge rapidly
beyond the weak field limit as $N$ increases as seen in Fig.
\ref{fig2}.

In the large-$N$ limit, the gaps scale to zero at $h_s=0$ as expected for the two
dimensional case.  It is also anticipated that the gaps of the ladders can be extrapolated
to the two-dimensional case beyond the weak-field region given their smooth change
and rapid convergence with increasing $N$.  To this end, we fit the field dependence of
the gaps using the analytical scaling function
\begin{equation}
\Delta=\Delta_{0}+\beta (h_s-h_{s_{0}})^{\gamma}\label{scaling},
\end{equation}
for both transverse\cite{Oshikawa1} and longitudinal
branch.\cite{Maslov, Ercolessi, Gliozzi} As expected, $\Delta_{0}$
and $h_{s_{0}}$ are zero for odd $N$ within the numerical accuracy
of the computation and scale toward zero monotonically and rapidly
for even $N$. Fitting to the DMRG results indicates that $h_{s_0}$
is already vanishingly small for $N$=4.  In Table I, we show the
direct fitting data for $N=1,3,4$ using Eq. (\ref{scaling}) and
the fitting curves are shown in Fig. \ref{fig2}.  Meanwhile, due
to the large gap at $h_s=0$, a good fitting using Eq.
(\ref{scaling}) cannot be achieved for $N$=2; instead, the scaling
parameters for $N$=2 are obtained from the extrapolation curves
obtained from fitting the results for $N$=1, 3, 4 (see below).

\begin{table}[h!]
\caption{Fitting coefficients and exponents for $N=1,2,3,4$ and
$\infty$.  Those for the two-leg ladder are taken from Fig.
\ref{fig3} at $N=2$ and $\beta$ and $\gamma$ for $N=\infty$ from
Fig. \ref{fig3} (see text). }
\begin{ruledtabular}
\begin{tabular}{c|ccccc}
N                        & 1      & 2     &    3    & 4    & $\infty$      \\
\hline
$\beta_{\cal L}^N     $  & 2.97   & 3.84  &    3.98 & 4.02  & $4.03\pm 0.09$    \\
$\gamma_{\cal L}^N   $  & 0.68   & 0.56  &    0.53 & 0.52  & $0.50\pm 0.01$   \\
$\beta_{\cal T}^N     $  & 1.88   & 2.10  &    2.16 & 2.19  & $2.27\pm 0.01$    \\
$\gamma_{\cal T}^N    $  & 0.61   & 0.56  &    0.54 & 0.53  &
$0.50\pm 0.01$
\end{tabular}
\end{ruledtabular}
\end{table}

\begin{figure}[h!]
\begin{center}
\includegraphics[width=6.60cm,angle=0]{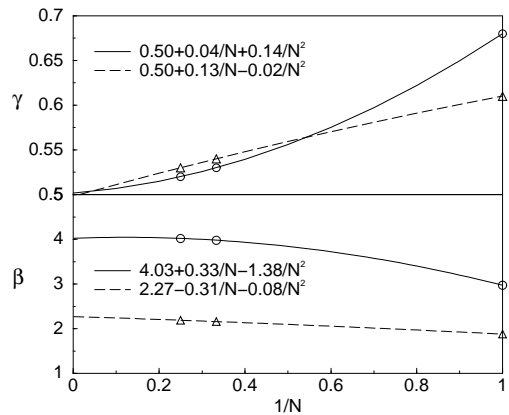}
\vspace{-0.5cm}
 \caption{The extrapolation of the scaling exponents
and coefficients of the field induced gaps from $N$-leg ladders
toward the square lattice ($N \rightarrow \infty$).} \label{fig3}
\end{center}
\end{figure}

We extrapolate the longitudinal and transverse gaps
$\gamma^N_{\cal L,T}$ and $\beta_{\cal L,T}^N$ toward the
large-$N$ (two dimensional) limit using the second order
polynomial fitting: $a(N)=a_{0}+a_1N^{-1}+a_2N^{-2}$, where $a$
denotes either $\gamma^{N}_{L,T}$ or $\beta_{L,T}^N$ given in
Table I with $N=1,3,4$. This fitting is illustrated in Fig.
\ref{fig3} and the extrapolated $\gamma^\infty_{\cal L,T}$ and
$\beta_{\cal L,T}^\infty$ are given in Table I.  From this
procedure, we obtain the longitudinal and transverse gaps for the
isotropic antiferromagnetic Heisenberg square lattice with a
staggered magnetic field
\begin{eqnarray}
&&\Delta_{\mathcal{L}}=(4.03\pm 0.09)h_s^{0.50\pm0.01},\\
&&\Delta_{\mathcal{T}}=(2.27\pm 0.01)h_s^{0.50\pm0.01}.\nonumber
\end{eqnarray}
Our numerical results suggest a common exponent $\gamma$ for both
branches. It is also interesting to note that a recent
field-theoretical study \cite{Sato} of coupled spin-1/2
antiferromagnetic Heisenberg chains with a leg-independent
staggered field also shows that the field induced gap scales with
the staggered magnetic field as $h_s^{1/2}$. Possible connections
and general implications of the common scaling behavior in these
different spin lattice models deserve further analytical
investigation.

Finally, we examine the scaling behavior of the midgap states in spin ladders with
a staggered magnetic field.  As shown in Fig. 4, midgap states exist in all the cases
studied and the scaling exponents $\gamma_{\cal L}^N$ and  $\gamma_{\cal T}^N$ for the
bulk gaps provide a good fit for the midgap states.  However, the scaling coefficients
$\beta_{\cal L}^N$ and $\beta_{\cal T}^N$ show an oscillatory pattern with increasing
$N$ (up to $N=4$ studied here).  It indicates that these coefficients have not properly
converged at $N=4$, probably due to a different (compared to the bulk excitations)
energy scale associated with the boundary edge excitations.\cite{Ng,Qin,to}  Consequently,
the question of whether the midgap states would persist in the large-$N$ limit remains
open at present.  Meanwhile, we observe that although the bulk low-energy behaviors of
the spin-1/2 two-leg ladder are similar to those of a spin-1 chain,\cite{SW1} their
boundary edge excitations and the field scaling behaviors are quite different.\cite{Wang00}
The spin-1 chain has a four-fold degenerate ground state that splits in a staggered field
and turns (partly) into the lowest midgap states.\cite{lou2}   The two-leg ladder has a
non-degenerate ground state and its transverse midgap states come from the
boundary edge excitations that originally overlap with the bulk excitation continuum.
However, when one adds other interactions to the standard two-leg ladder\cite{SW1,Wang00},
the resulting low-energy properties, ie. gap and midgaps, in the Haldane phase
can be the same as the $S=1$ chain\cite{lou2}.
\begin{figure}
\begin{center}
\includegraphics[width=7.5cm]{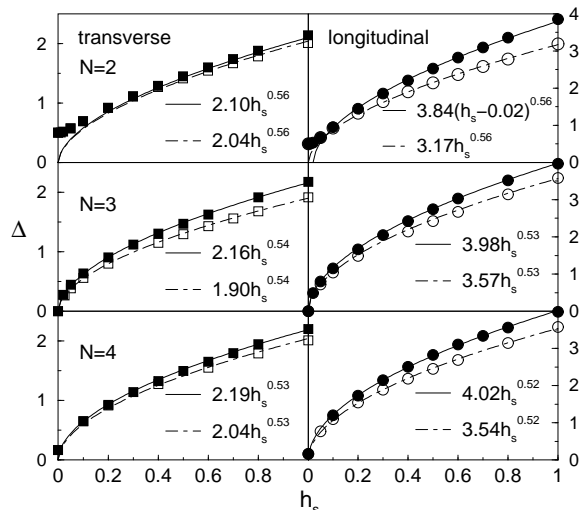}
\vspace{-0.5cm}
 \caption{The scaling fit for the transverse and
longitudinal gaps and the midgap states for the ladders (top to
bottom panels: $N$=2$-$4). The filled and open symbols represent
the gap and midgap states, respectively.  The exponents and
coefficients for $N$=2 are taken from Fig. \ref{fig3}.}
\label{fig4}
\end{center}
\end{figure}

In summary, we have carried out a systematic study of the
low-energy excitations of spin-1/2 antiferromagnetic Heisenberg
chain and ladders in a staggered magnetic field and obtained the
field scaling behavior of both longitudinal and transverse gaps. A
modified boundary scheme has been devised to extract the bulk
excitation behavior using the more accurate open boundary option
in the DMRG calculations. The calculated bulk gaps converge
rapidly with increasing number of legs above weak field regime; it
allows a reliable extrapolation to obtain the field scaling
behavior of the gaps for the isotropic square lattice.  We also
examined the midgap states in the spin ladders induced by the
topological edge effect.  The midgap states share the same scaling
exponents with the bulk gaps but the scaling coefficients show an
oscillatory pattern with increasing number of legs (up to $N$=4).
As a result, the overall scaling behavior of the midgap states in
the large-$N$ limit needs further investigation.

This work was supported by the National Basic Research Program
under the Grant 2005CB32170X and the NSFC Under Grant No. 10425417
and 90203006, and by the U. S. Department of Energy under
Cooperative Agreement DE-FC52-01NV14049 at UNLV.  X.W.
acknowledges the support during his visit at UNLV. The
computations were performed on ICTS-HPSC45.

\end{document}